\documentclass[a4paper,12pt,twoside]{article}

\usepackage[dvips]{graphicx}
\usepackage{geometry}
\usepackage[centertags]{amsmath}
\usepackage{amsfonts}
\usepackage{amssymb}
\usepackage{amsthm}
\usepackage{latexsym}
\usepackage{psfrag}
\usepackage{fancyhdr}

\renewcommand{\vec}[1]{\ensuremath{\boldsymbol{\mathrm{#1}}}}

\newcommand{\ArXivNo}{physics/0412040}

\newcommand{\XXXSize}{{\fontsize{12}{12}\selectfont\fbox{\textbf{\ArXivNo}}}}
\newcommand{\XXXTitle}{\hfill\XXXSize\newline\vskip 0.4cm}

\begin{document}
\pagestyle{fancy}

\sloppy

\title{\XXXTitle\textbf{Projection of relativistically moving
objects on a two-dimensional plane, the `train' paradox and the
visibility of the Lorentz contraction}%
\thanks{Published in \textit{European~Journal~of~Physics},
Vol.~\textbf{23}, No.~2, (2002) pp.~103--110. \newline %
[\url{http://dx.doi.org/10.1088/0143-0807/23/2/302}] %
}}

\author{\textsc{E.~B.~Manoukian}\thanks{E-mail: \texttt{edouard@ccs.sut.ac.th}}
\ and \ \textsc{S.~Sukkhasena} \\
{School of Physics, \ Suranaree University of Technology} \\
\ Nakhon Ratchasima, 30000, Thailand }
\date{} \maketitle

\begin{abstract}
Although many papers have appeared on the theory of photographing
relativistically moving objects, pioneered by the classic work of
Penrose and Terrell, three problems remain outstanding\@. \
\ref{Item1}.~There does not seem to exist a general formula which
gives the projection of a relativistically moving object,
applicable to any object no matter how complicated, on a
two-dimensional plane in conformity with Terrell's observation\@.
\    \ref{Item2}.~No resolution seems to have been provided for
the associated so-called `train' paradox\@. \      \ref{Item3}.~No
analytical demonstration seems to have been offered on how the
Lorentz contraction may be actually detected on a photograph\@. \
This paper addresses all of these three problems\@. \     The
analysis does not require any more than
trigonometry and elementary differentiation\@. \\
\end{abstract}

\section{Introduction}\label{Sec1}

Since the early classic work of Penrose~\cite{Penrose_1959},
Terrell~\cite{Terrell_1959} and Weisskopf~\cite{Weisskopf_1960} on
the appearance of relativistically moving objects several
refinements and extensions of the above work have been carried out
(see, e.g., \cite{Yngstrom_1962, Scott_1965, McGill_1968,
Scott_1970, Mathews_1972, Hickey_1979, Burke_1991,
Howard_1995})\@. \   Three problems seem, however, to remain
outstanding\@.

\makeatletter
\renewcommand{\theenumi}{\arabic{enumi}}
\renewcommand{\labelenumi}{\theenumi.}
\renewcommand{\theenumii}{\roman{enumii}}
\renewcommand{\labelenumii}{(\theenumii)}
\renewcommand{\p@enumii}{}
\makeatother
\begin{enumerate}
\item\label{Item1}  There does not seem to be in the literature
a general formula which provides the projection of a
relativistically moving object, applicable to \emph{any} object no
matter how complicated, on a two-dimensional plane in conformity
with Terrell's observation\@. \    Terrell's observation is that
different points on the object must `emit' light at different
times in order to reach an observation point simultaneously\@.

\item\label{Item2}  No resolution seems to have been offered of one of
the most puzzling aspects in the above investigations referred to
as the `train' paradox\@. \     No much attention has been given
to this in the literature\@. \    This paradox has its roots in
the early work of Terrell~\cite{Terrell_1959} and
Weisskopf~\cite{Weisskopf_1960} and was emphasized by Mathews and
Lakshmanan~\cite{Mathews_1972} almost 30 years ago\@. \     In its
simplest terms, the latter paradox arises in the following
manner\@. \    One often reads in the above papers that an object
appears to be rotated when in relative motion to an observation
frame due to the fact that different points on the object must
`emit' light at different times in order to reach an observation
point simultaneously\@. \     Such an inference seems to indicate
that a rectangular block, for example, sliding on (smooth) rails
and the edges of its lower side in contact with them, appears off
them due to the relative motion with the rails stationary relative
to the observer, and hence the paradox\@. \     The same reasoning
may be applied, as shown below, to an object with a horizontal
flat top touching a `smooth' flat horizontal plane\@. \     Again
this would seem to imply, in particular cases, that one end of the
object had miraculously broken and gone through the flat plane due
to the relative motion\@.

\item\label{Item3}  No analytical demonstration seems to have been
given on how the Lorentz contraction may be detected on a
photograph\@. \    The key point that we discover is that, due to
Terrell's observation, the portion of a moving ruler to the left
of the observer seems elongated, while the portion of the ruler to
the right of the observer seems to be shrunk\@. \    These effects
work, respectively, against and with the Lorentz contraction, in
general masking it\@. \     Accordingly, we may locate a critical
point on the ruler such that the Lorentz contraction is visible
for a small interval on the ruler around this point\@.

The word `appears' or the statement `appears as on a photograph'
have caused some confusion over the years\@. \     Any such
wordings necessarily involve assumptions used in one's analysis\@.
\    To be precise, the latter are meant in the following manner,
as working hypotheses, in the present investigation and are based
on taking into account these three points~:

\begin{enumerate}
\item\label{Item3i}  Terrell's observation that different points
on an object must `emit' light at different times in order to
reach an observation point simultaneously\@. \    There is an
inherit limitation in considering a point observation site\@. \
The removal of such a restriction is certainly a formidable task,
which will not be attempted in this paper\@.

\item\label{Item3ii}  The Lorentz transformations.

\item\label{Item3iii}  The piercing by these light rays of
an appropriate two-dimensional plane in the observation frame\@. \\
\end{enumerate}
\end{enumerate}

In section~\ref{Sec2}, we provide a complete and elementary
derivation of the explicit (nonlinear) transformations arising
from the application of the three points
(\ref{Item3i})--(\ref{Item3iii}), which may be directly applied to
\emph{any} object no matter how complicated\@. \    The formulae
obtained are easily accessible to students and sufficiently
precise to illustrate faithfully the main features of the Terrell
effect\@. \    These transformations are appropriately referred to
as nonlinear Terrell transformations\@. \     The closest
investigation to these transformations was given by
Hickey~\cite{Hickey_1979}, who, however, applied methods of
mapping out the tangents to points on an object\@. \    The latter
also provides no room for resolving the `train' paradox\@. \
Although the demonstration of the absence of a paradox seems
non-trivial, we provide in section~\ref{Sec3} a rather
\emph{elementary} resolution of the `train' paradox\@. \    The
visibility of the Lorentz contraction is studied in
section~\ref{Sec4}, where a critical point on a moving ruler is
singled out around which the Lorentz contraction is visible for a
partitioning of the ruler around this point\@. \
Section~\ref{Sec5} deals with our conclusions\@. \\

\setlength{\fboxsep}{0.02\textwidth}
\begin{figure}[!htb]
  \centering
  \fbox{ \begin{minipage}{0.94\textwidth}
  \centering
  \psfrag{O}{$\mathrm{O}$} \psfrag{OO}{$\overline{\mathrm{O}}$}
  \psfrag{x}{$x$} \psfrag{y}{$y$}
  \psfrag{z}[][]{$z$} \psfrag{h}{$(0,0,h)$}
  \psfrag{U}{$U$} \psfrag{V}{$V$}
  \psfrag{n}{$\vec{\hat{n}}$} \psfrag{d}{$d$}
  \psfrag{UV}[][]{$(U,V)$} \psfrag{xx}{$(x_{0},y_{0},h)$}
  \psfrag{xy}{$(x,y,z)$} \psfrag{xh}{$(x,y,h)$}
  \includegraphics[width=12cm]{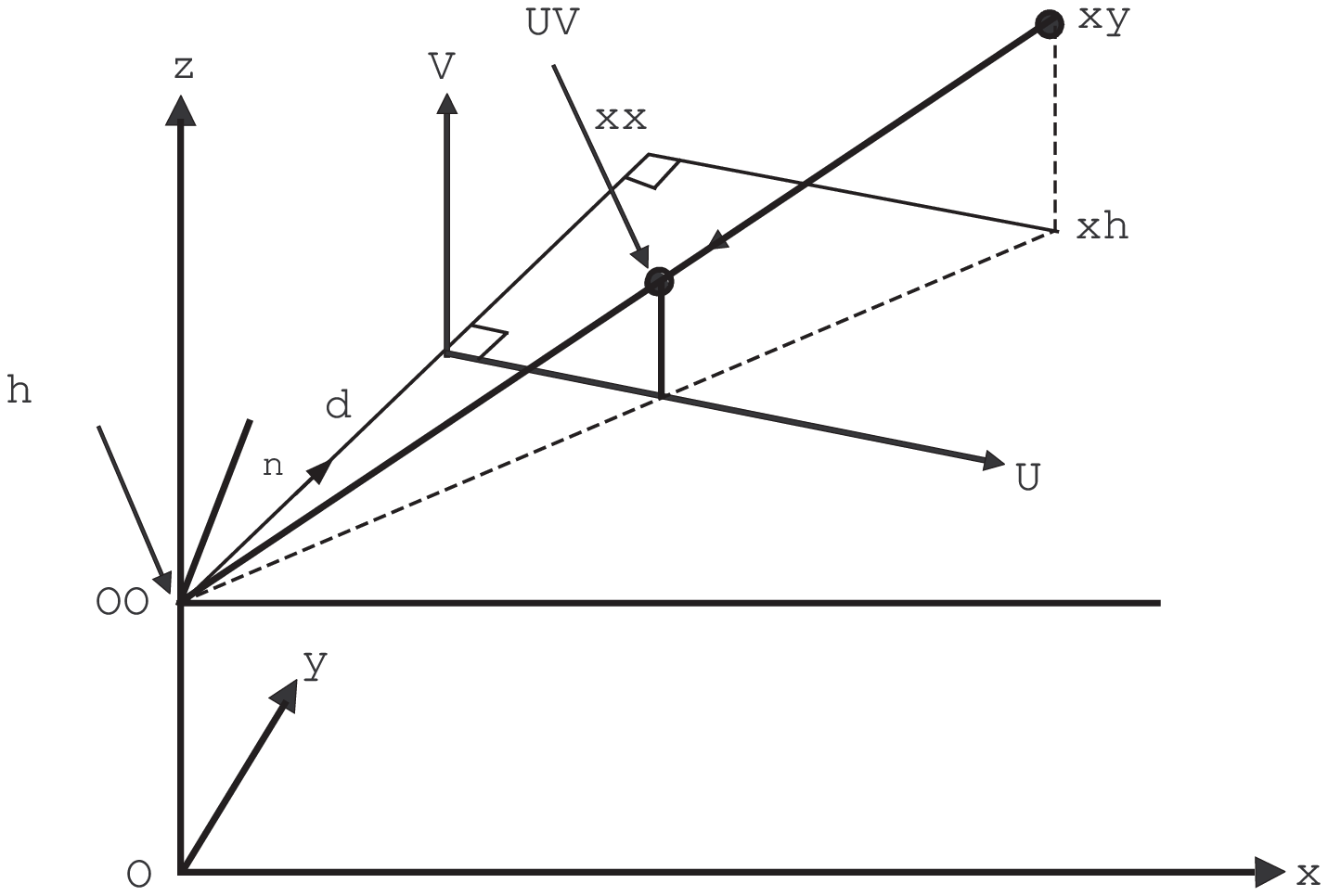}
  \caption[]{The observation coordinate system\@. \ The observer is
  at $\overline{\mathrm{O}}$ at a height $h$ above the origin\@. \
  The $U$-$V$ plane is specified by a unit vector $\vec{\hat{n}}$
  chosen parallel to the $x$-$y$ plane\@. \    The origin of the
  $U$-$V$ plane is at a distance $d$ from $\overline{\mathrm{O}}$\@. \
  A light ray from $(x,y,z)$, a given point on the object in relative
  motion as labelled in the observation frame, on its way to
  $\overline{\mathrm{O}}$ pierces the $U$-$V$ plane at $(U,V)$\@.}
  \label{Fig01}
  \end{minipage} }
\end{figure}
%\clearpage

\section{Nonlinear Terrell transformations}\label{Sec2}

Consider the relative motion of the proper inertial frame
$\mathrm{F}'$ of an object to be along the $x$-axis of an
observation frame $\mathrm{F}$ with speed $v$\@. \    Let
$(x',y',z')$, $(x,y,z)$ denote, respectively, the labellings of an
arbitrary point on the object in the corresponding $\mathrm{F}'$
and $\mathrm{F}$ frames\@. \    The observation point
$\overline{\mathrm{O}}$ is at a height $h$, along the $z$-axis
(see figure~\ref{Fig01}) above the origin $\mathrm{O}$ of the
$\mathrm{F}$-frame\@. \     When the origins $\mathrm{O}$ and
$\mathrm{O}'$, at $t=0$ and $t'=0$, of $\mathrm{F}$ and
$\mathrm{F}'$ frames coincide and the observer at
$\overline{\mathrm{O}}$ sees this happening he takes a snap shot
of the object\@. \     Since the observer at
$\overline{\mathrm{O}}$ sees the origin coincide only at a later
time equal to $h/c$, the time that light was `emitted' from
$(x,y,z)$ to reach $\overline{\mathrm{O}}$ is given by
\begin{equation}\label{Eqn01}
  t = -\frac{\sqrt{x^{2}+y^{2}+(z-h)^{2}}}{c}+\frac{h}{c}.
\end{equation}
We are interested in all the light rays from the object which
reach point $\overline{\mathrm{O}}$ simultaneously\@. \     The
light ray from a point $(x,y,z)$ on the object will pierce a
$U$-$V$ plane specified by a unit vector $\vec{\hat{n}}$ as shown
in figure~\ref{Fig01} perpendicular to the former\@. \     Here
$U$ and $V$ define coordinate axes on the two-dimensional plane on
which the object is projected\@. \     A point $(x',y',z')$ on the
object, as labelled in its rest frame $\mathrm{F}'$, will be
mapped into a point $(U,V)$ on the two-dimensional plane\@. \  $d$
denotes the distance from $\overline{\mathrm{O}}$ (the observer)
to the origin of the $U$–$V$ plane\@. \     For concreteness
$\vec{\hat{n}}$, emerging from $\overline{\mathrm{O}}$ and
defining what is called the optic axis, was chosen to be parallel
to the $x$-$y$ plane\@. \     The former, in turn, is defined by
choosing a point $(x_{0},y_{0},h)$ as shown in figure~\ref{Fig01},
and hence may be written as
\begin{equation}\label{Eqn02}
  \vec{\hat{n}} = \left(\frac{x_{0}}{\sqrt{x_{0}^{2}+y_{0}^{2}}},
  \frac{y_{0}}{\sqrt{x_{0}^{2}+y_{0}^{2}}},0\right)
  \equiv{} \left(n_{1},n_{2},0\big.\right).
\end{equation} \\

From figure~\ref{Fig01},
\begin{equation}\label{Eqn03}
  (x-x_{0})^{2}+(y-y_{0})^{2}+x_{0}^{2}+y_{0}^{2} = x^{2}+y^{2}
\end{equation}
or
\begin{equation}\label{Eqn04}
  x_{0}^{2}+y_{0}^{2} = (xn_{1}+yn_{2})^{2}
\end{equation}
and
\begin{equation}\label{Eqn05}
  \frac{U^{2}}{d^{2}} = \frac{(x-x_{0})^{2}+(y-y_{0})^{2}}{x_{0}^{2}+y_{0}^{2}}
\end{equation}
giving the solution
\begin{equation}\label{Eqn06}
  U = d\,\frac{(xn_{2}-yn_{1})}{(xn_{1}+yn_{2})}.
\end{equation}
Also
\begin{equation}\label{Eqn07}
  \frac{V}{(z-h)} = \frac{\sqrt{d^{2}+U^{2}}}{\sqrt{x^{2}+y^{2}}}
\end{equation}
giving the solution
\begin{equation}\label{Eqn08}
  V = d\,\frac{(z-h)}{(xn_{1}+yn_{2})}.
\end{equation} \\

The transformations (\ref{Eqn06}) and (\ref{Eqn08}), as shown
above, follow from the examination of figure~\ref{Fig01}\@. \
These transformations will now allow us to find the mapping of a
point $(x',y',z')$ on the object, as labelled in its rest frame
$\mathrm{F}'$, into the $U$-$V$ plane\@. \     To this end, from
the Lorentz transformations $x'=\gamma(x-vt)$, $y'=y$, $z'=z$,
using equation (\ref{Eqn01}), we may solve for $x$ and substitute
the latter in (\ref{Eqn06}) and (\ref{Eqn08}) to obtain the
transformations we are seeking, where we set $v/c=\beta$,
\begin{align}
  U &= d\,\frac{\gamma\left[(x'+\gamma\beta{}h)-\beta\sqrt{(x'+\gamma\beta{}h)^{2}
  +{y'}^{2}(z'-h)^{2}}\Big.\right]n_{2}-y'n_{1}}{\gamma\left[(x'+\gamma\beta{}h)
  -\beta\sqrt{(x'+\gamma\beta{}h)^{2}+{y'}^{2}(z'-h)^{2}}\Big.\right]n_{1}+y'n_{2}}
  \label{Eqn09} \\
  V &= d\,\frac{(z'-h)}{\gamma\left[(x'+\gamma\beta{}h)
  -\beta\sqrt{(x'+\gamma\beta{}h)^{2}+{y'}^{2}(z'-h)^{2}}\Big.\right]n_{1}+y'n_{2}}
  \label{Eqn10}
\end{align}
where $\gamma=(1-\beta^{2})^{-1/2}$ and $(x',y',z')$ denotes any
given point on the object in its \emph{rest} frame\@. \     Unlike
the Lorentz transformations $(t,x,y,z)\to{}(t',x',y',z')$ for a
time-space point, the transformations in (\ref{Eqn09}) and
(\ref{Eqn10}), $(x',y',z')\to{}(U,V)$, are obviously nonlinear\@.
\      The latter may be appropriately referred to as nonlinear
Terrell transformations\@. \\

\begin{figure}[!hb]
  \centering
  \fbox{ \begin{minipage}{0.94\textwidth}
  \centering
  \includegraphics[width=0.93\textwidth]{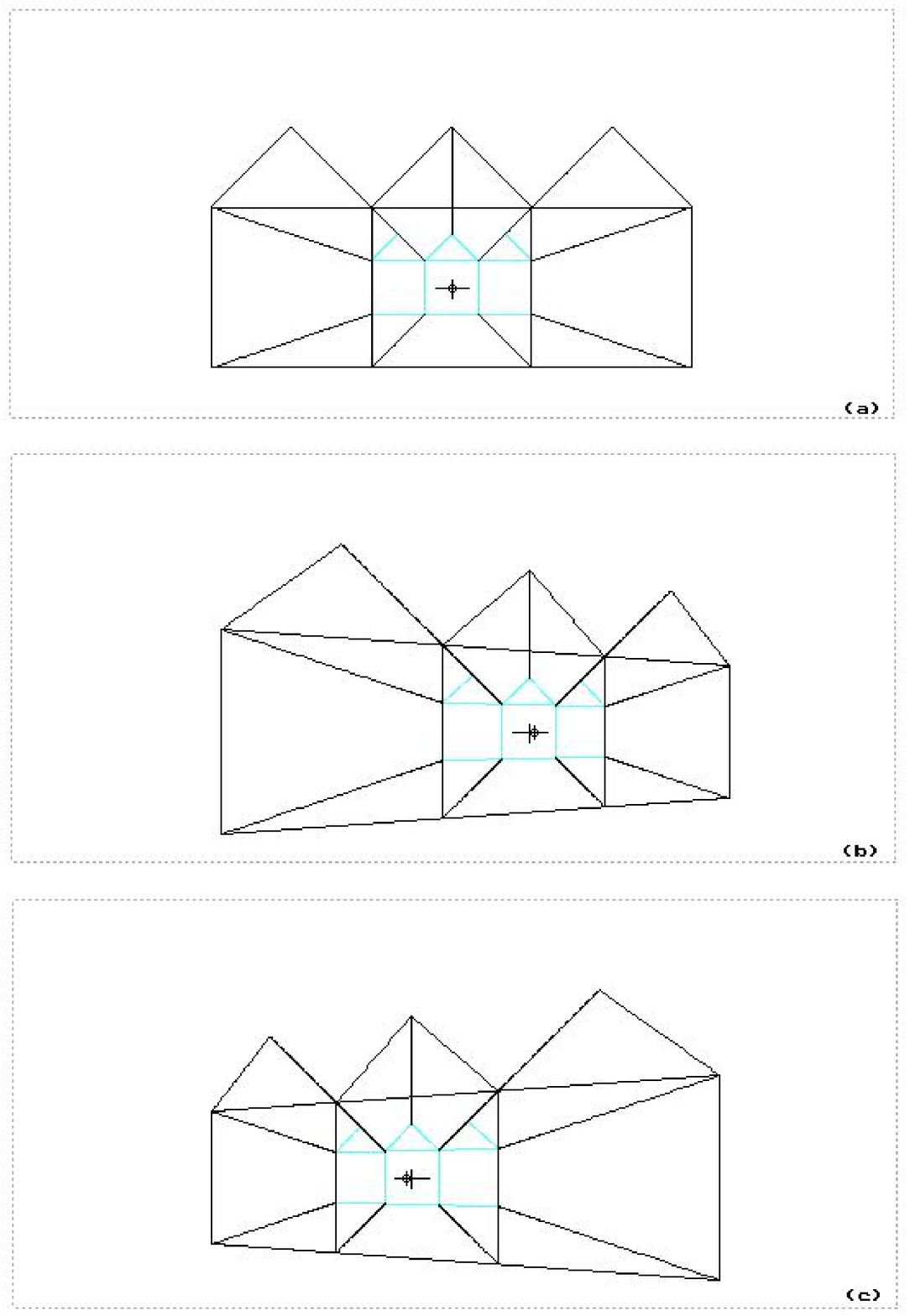}
  \caption[]{Projection on the $U$-$V$ plane for $\beta=0$,
  (a)~$\vec{\hat{n}}=\left(0,1,0\big.\right)$, \
  (b)~$\vec{\hat{n}}=\left(0.0712,0.9975,0\big.\right)$, \
  (c)~$\vec{\hat{n}}=\left(-0.0712,0.9975,0\big.\right)$\@. \
  The crossed line denotes a midpoint of the houses\@. \
  The crossed small circle denotes the centre of the $U$-$V$ plane\@.}
  \label{Fig02}
  \end{minipage} }
\end{figure}

\begin{figure}[!ht]
  \centering
  \fbox{ \begin{minipage}{0.94\textwidth}
  \centering
  \includegraphics[width=0.93\textwidth]{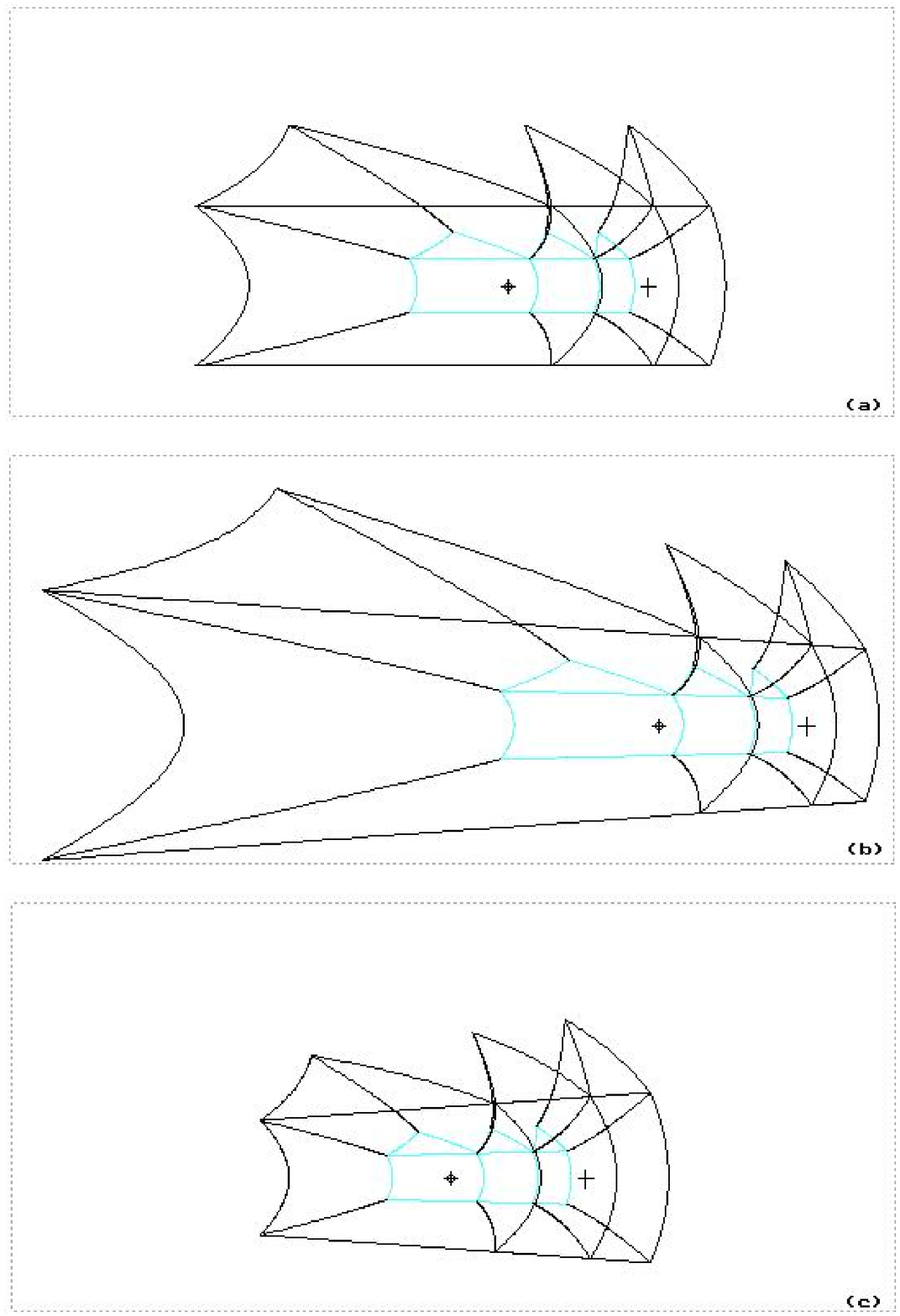}
  \caption[]{Projection on the $U$-$V$ plane for $\beta=0.9$ with
  $\vec{\hat{n}}$ in (a), (b), (c), respectively, as in
  figure~\ref{Fig02}\@. \     The observation frame moves to
  the left with speed $\beta{}c$\@.}
  \label{Fig03}
  \end{minipage} }
\end{figure}

An explicit application of the transformations (\ref{Eqn09}) and
(\ref{Eqn10}) is given in figures~\ref{Fig02} and \ref{Fig03},
respectively, for $\beta=0$ and $0.9$ for various directions of
the optic axis specified by the unit vector $\vec{\hat{n}}$\@. \
Pertinent remarks concerning these figures will be made in
section~\ref{Sec5}\@. \\

\clearpage

\clearpage

\begin{figure}[!htb]
  \centering
  \fbox{ \begin{minipage}{0.94\textwidth}
  \centering
  \psfrag{xx}{$x'$} \psfrag{yy}{$y'$}
  \psfrag{zz}[][]{$z'$} \psfrag{a}{$a$}
  \psfrag{P}[][]{$\mathrm{P}:\,(a,y',z')$}
  \includegraphics[width=7cm]{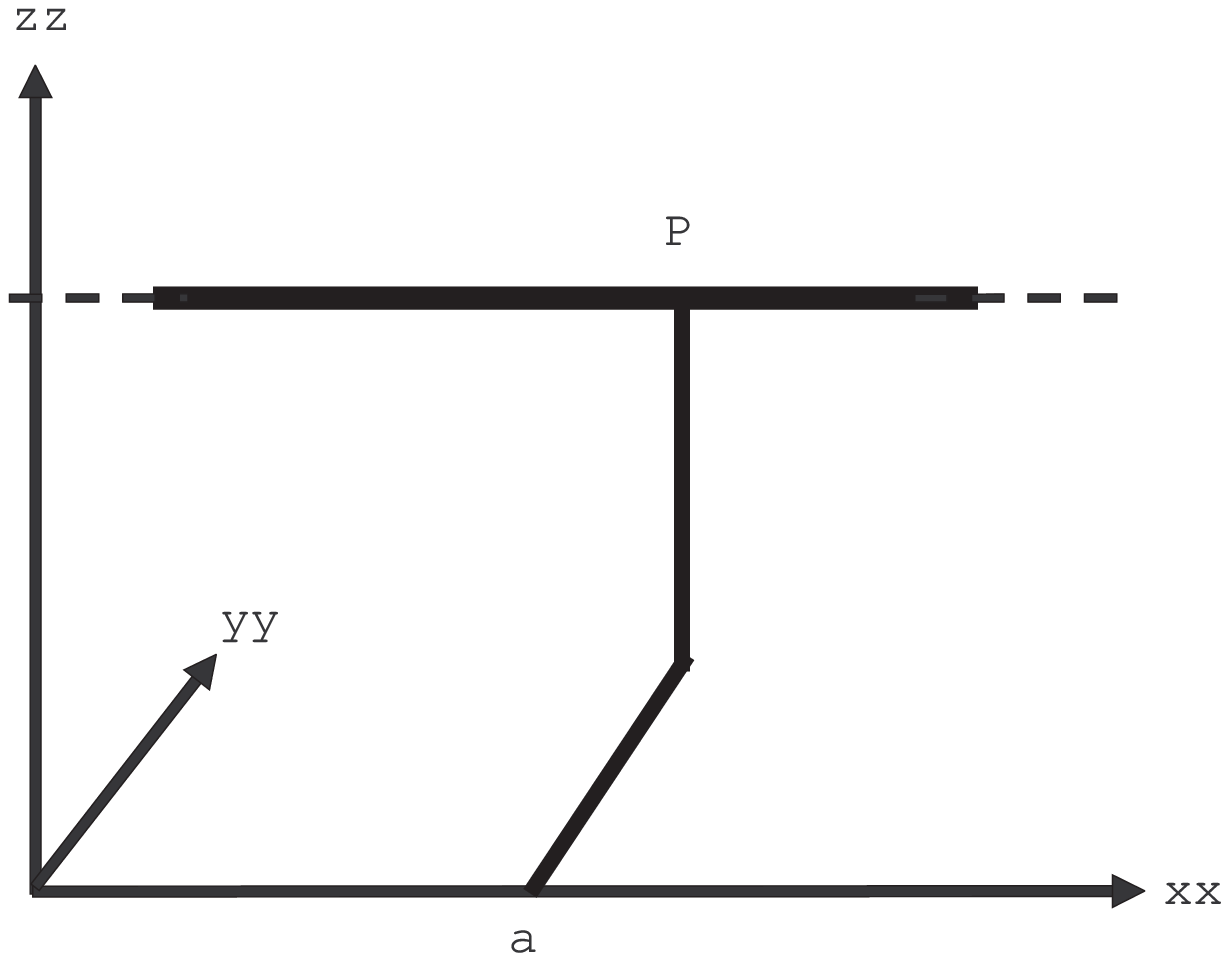}
  \caption[]{A given horizontal line parallel to the $x'$-axis\@. \
  The point $(a,y',z')$ on the object, for $\beta=0$, touches
  the line at $\mathrm{P}$\@.}
  \label{Fig04}
  \end{minipage} }
\end{figure}

\section{Resolution of the `train' paradox}\label{Sec3}

Let first $\beta=0$\@. \     Consider a horizontal line parallel
to the $x'$-axis, that is, take $y'$ and $z'$ as fixed\@. \
Suppose that the object in question touches this line at some
point, say, $(a,y',z')$\@. \     $x'$ takes on arbitrary values
along the line with $x'=a$ corresponding to the point of contact
just mentioned (see figure~\ref{Fig04})\@. \     That is, $a$ is
the $x'$ coordinate of a point of contact with the horizontal line
(for $\beta=0$) and also provides a reference point from which
other points will be defined as we shall now see\@. \     From
(\ref{Eqn06}) and (\ref{Eqn08})
\begin{equation}\label{Eqn11}
  U_{a} = d\,\frac{(an_{2}-y'n_{1})}{(an_{1}+y'n_{2})},\qquad
  V_{a} = d\,\frac{(z'-h)}{(an_{1}+y'n_{2})}
\end{equation}
giving
\begin{align}
  U-U_{a} &= d\,\frac{y'(x'-a)}{(an_{1}+y'n_{2})(x'n_{1}+y'n_{2})}
  \label{Eqn12} \\
  V-V_{a} &= -d\,\frac{(z'-h)n_{1}(z'-a)}{(an_{1}+y'n_{2})(x'n_{1}+y'n_{2})}.
  \label{Eqn13}
\end{align}
That is,
\begin{equation}\label{Eqn14}
  V-V_{a} = -\frac{(z'-h)}{y'}\,n_{1}(U-U_{a})
\end{equation}
or
\begin{equation}\label{Eqn15}
  V = \frac{(z'-h)}{y'}\left[-n_{1}U+n_{2}d\big.\right].
\end{equation}
Hence the above horizontal line is mapped into a straight line in
the $U$-$V$ plane\@. \     We recall that (\ref{Eqn15}) was
derived for $\beta=0$\@. \     For $\beta\neq{}0$, it is readily
checked that $U$ and $V$ given, respectively, in (\ref{Eqn09}) and
(\ref{Eqn10}) fall on the same line for $x'=a$\@. \     That is,
for $\beta\neq{}0$, $U$ and $V$ also satisfy equation
(\ref{Eqn15})\@. \     \emph{Hence any point on the object which
for $\beta=0$ lies on the straight line \emph{(\ref{Eqn15})} must
also lie on the same line for $\beta\neq{}0$}\@. \     This
provides the resolution of the `train' paradox\@. \\

For example, refer to figure~\ref{Fig02}(b) (and (a)) and consider
an imaginary line joining the tips of the three roofs as drawn in
the observation frame\@. \     Now consider the corresponding case
for $\beta=0.9$ in figure~\ref{Fig03}(b) with the houses pulled,
so to speak, to slide along this line relative to a stationary
observer\@. \     Here one has the impression that the tip of the
roof of the first house on the left has cut through and passed
through this line\@. \     The above demonstration shows that the
tips of the roofs of all the three houses remain \emph{always in
contact} for $\beta\neq{}0$ as well with the straight line but at
different points for $\beta=0$ and $\beta\neq{}0$ due to the
relative motion\@. \\

\section{Visibility of the Lorentz contraction}\label{Sec4}

\begin{figure}[!htb]
  \centering
  \fbox{ \begin{minipage}{0.94\textwidth}
  \centering
  \psfrag{1}{$1$} \psfrag{2}{$2$}
  \psfrag{11}{$1'$} \psfrag{22}{$2'$}
  \psfrag{3}{$3$} \psfrag{4}{$4$}
  \psfrag{33}{$3'$} \psfrag{44}{$4'$}
  \psfrag{O}[][]{Observer}
  \includegraphics[width=10cm]{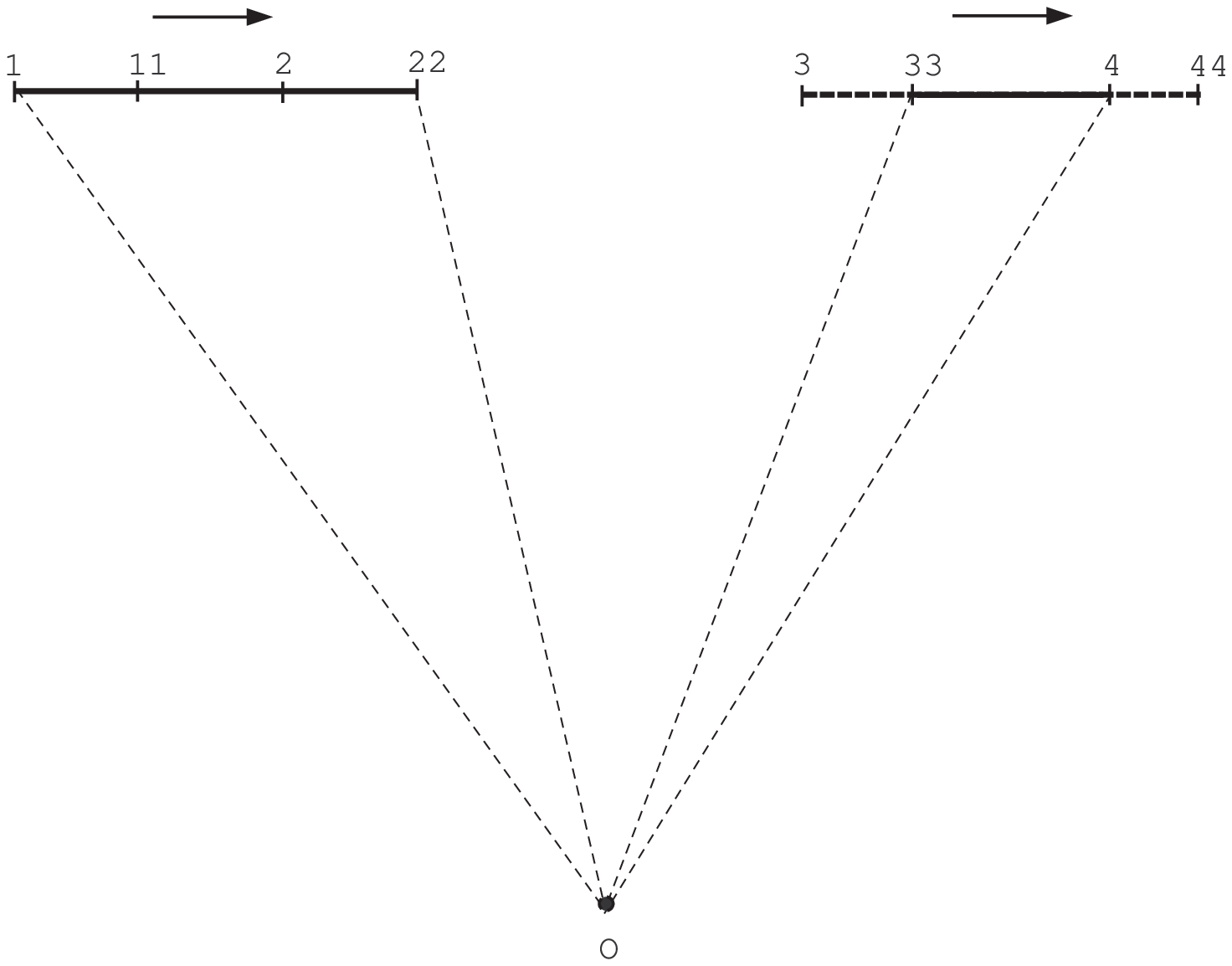}
  \caption[]{For light `emitted' from both ends of the moving
  ruler on the left to reach the observer simultaneously,
  the ruler labelled by $1$, $2$ in its proper frame is
  effectively defined by $1$, $2'$ to the observer and seems
  to be elongated\@. \     Similarly, the ruler on the right,
  which is labelled by $3$, $4$ in its proper frame, is
  effectively defined by $3'$, $4$ to the observer and
  seems to be shrunk\@.}
  \label{Fig05}
  \end{minipage} }
\end{figure}

Consider two rulers of equal proper lengths each moving to the
right with speed $v$, with one approaching and one receding from
the observer (see figure~\ref{Fig05})\@. \     The end points of
the ruler on the left are labelled by $1$, $2$, while the ones on
the right by $3$, $4$ (see figure~\ref{Fig05})\@. \      In
reference to the ruler on the left (approaching the observer), in
order that light `emitted' from the end points $1$ and $2$ reach
the observer simultaneously, light should be `emitted' from point
$1$ first\@. \      In the meantime, the ruler moves to the right
and the end point $2$ reaches some point $2'$ and `emits' light to
reach the observer at the same time as light that was `emitted'
from end point $1$ earlier\@. \      Accordingly, the ruler seems
to be elongated with end points effectively defined by $1$ and
$2'$\@. \      On the other hand, the same reasoning applied to
the ruler on the right (receding from the observer) shows that the
ruler seems to be shrunk with end points effectively defined by
$3'$ and $4$\@. \      These facts have nothing to do with the
Lorentz transformations or the Lorentz contraction, but they would
contribute in forming the final projected image on the
two-dimensional plane, and would, in general, mask the visibility
of the Lorentz contraction, which is of different nature\@. \\

The old fundamental and critical question now comes to haunt us\@.
\      Can we photograph the Lorentz contraction\@? \        To
answer this question, consider a ruler moving to the right with
speed $\beta{}c$ such that its lower edge is at $z'=0$, and
$y'\neq{}0$ is arbitrary but has some fixed value\@. \      We
consider the ruler to have a portion of it to the right of the
observer and a portion to the left of it\@. \       We set the
observer at the origin $\mathrm{O}$ of his coordinate system, i.e.
set $h=0$\@. \      We also choose the optic axis to be specified
by the unit vector $\vec{\hat{n}}=\left(0,1,0\big.\right)$\@. \
Then
\begin{equation}\label{Eqn16}
  U = \frac{d}{y'}\,\gamma\left[x'-\beta\sqrt{{x'}^{2}+{y'}^{2}}\Big.\right]
\end{equation}
($V=0$) corresponding to the lower edge of the ruler\@. \\

Suppose that the ruler is partitioned with small increments
$\Delta{}x'$\@. \     Consider a small change $\Delta{}U$ in
(\ref{Eqn16}) along $x'$-axis, that is along the ruler,
corresponding to a small increment $\Delta{}x'$\@. \      By
elementary differentiation, (\ref{Eqn16}) gives
\begin{equation}\label{Eqn17}
  \Delta{}U = \frac{d}{y'}\,\gamma\left[1-\beta\frac{x'}{\sqrt{{x'}^{2}+{y'}^{2}}}
  \right]\Delta{}x'
\end{equation}
which may be conveniently rewritten in the equivalent form
\begin{equation}\label{Eqn18}
  \Delta{}U = \frac{d}{y'}\left[\frac{\Delta{}x'}{\gamma}+\gamma\beta\left(
  \beta-\frac{x'}{\sqrt{{x'}^{2}+{y'}^{2}}}\right)\Delta{}x'\right].
\end{equation}
Since $\left|x'/\sqrt{{x'}^{2}+{y'}^{2}}\right|<1$, we may infer
that for any given $\beta<1$, around the critical point defined by
\begin{equation}\label{Eqn19}
  \frac{x'}{\sqrt{{x'}^{2}+{y'}^{2}}} = \beta
\end{equation}
we have
\begin{equation}\label{Eqn20}
  \Delta{}U = \frac{d}{y'}\frac{\Delta{}x'}{\gamma} \qquad{}
  \left(\left.\Delta{}U\Big.\right|_{\beta=0} = \frac{d}{y'}\,\Delta{}x'\right)
\end{equation}
which apart from the trivial constant scaling factor $d/y'$, is
the famous Lorentz contraction formula\@. \     \emph{That is,
around a certain point of the ruler, the Lorentz contraction is
visible} in the $U$-$V$ plane on the two-dimensional surface for a
small increment of proper length $\Delta{}x'$ drawn around this
critical point\@. \\

\begin{figure}[!htb]
  \centering
  \fbox{ \begin{minipage}{0.94\textwidth}
  \centering
  \psfrag{xx}{$x'$} \psfrag{yy}{$y'$}
  \psfrag{ac}{$\cos^{-1}(\beta)$} \psfrag{sqr}{$\sqrt{{x'}^{2}+{y'}^{2}}$}
  \includegraphics[width=8cm]{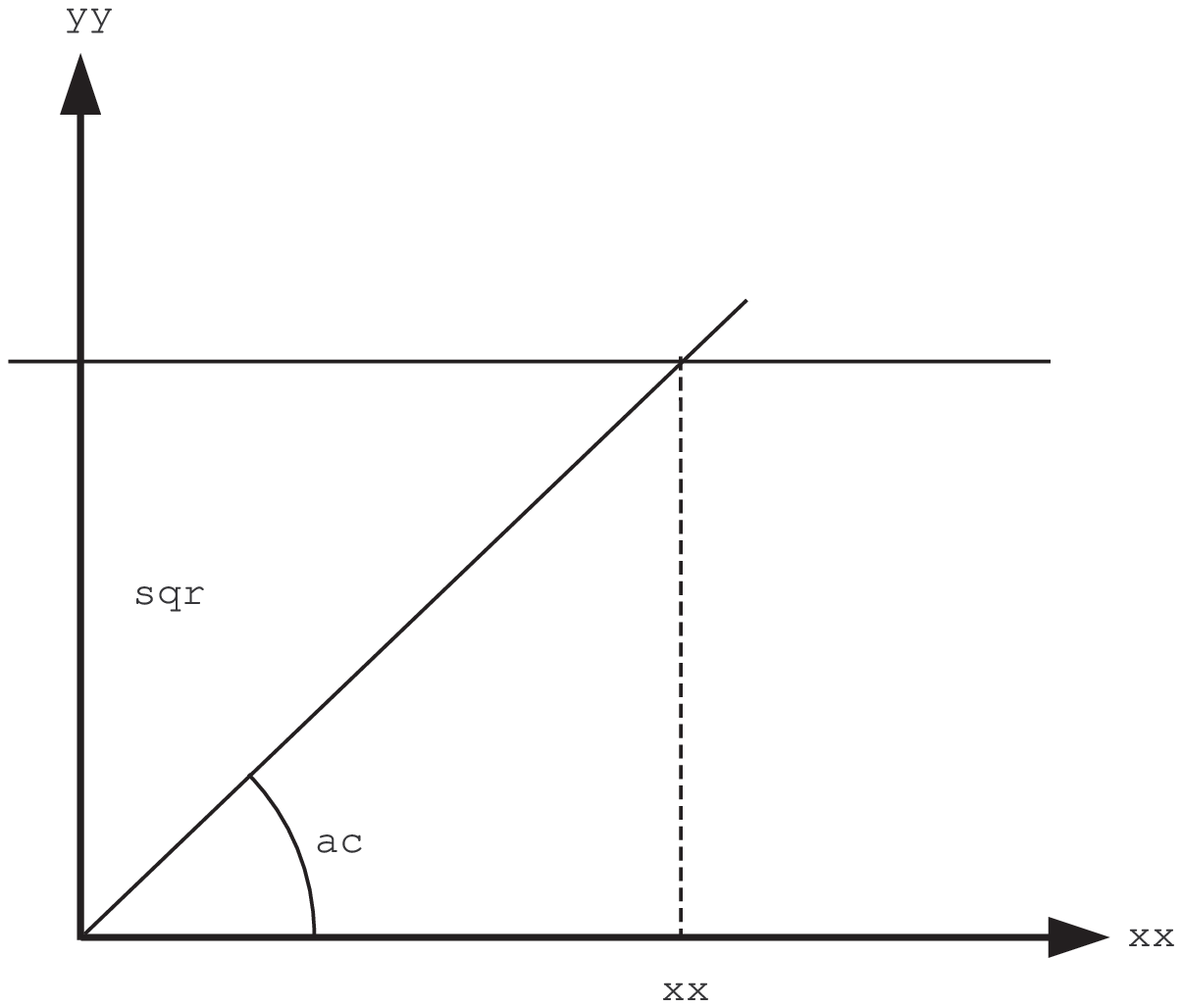}
  \caption[]{Line making an angle $\theta=\cos^{-1}(\beta)$ with
  the $x'$-axis in the ruler's proper frame\@. \     This line
  crosses the ruler at the critical point $x'$ about which the
  Lorentz contraction is visible in the $U$-$V$ plane for a small
  interval $\Delta{}x'$ about $x'$\@.}
  \label{Fig06}
  \end{minipage} }
\end{figure}

To find the above critical point in question on the ruler, draw a
line making an angle $\theta=\cos^{-1}(\beta)$ with the $x'$-axis
(see figure~\ref{Fig06}) before setting the ruler to move with a
given pre-assigned speed $\beta{}c$\@. \      This line will cross
a point on the lower edge of the ruler defining the critical point
in question\@. \\

\section{Conclusions}\label{Sec5}

That there is contraction on the right-hand side of the houses
(figure~\ref{Fig03}) and relative elongation on their left-hand
sides is clear when referring to figure~\ref{Fig05} and is due to
the fact that light `emitted' from the houses from different parts
must be `emitted' at different times to reach the observation
point $\overline{\mathrm{O}}$ simultaneously\@. \      We have
also seen that all the lines that are parallel to the $x'$-axis
remain straight lines in the $U$-$V$ plane for all $\beta$, and
this is the content of the resolution of the `train' paradox given
in section~\ref{Sec3}\@. \     It is readily checked from
(\ref{Eqn09}) and (\ref{Eqn10}), that any lines parallel to the
$y'$ or the $z'$ axes for $\beta=0$ become curved in the $U$-$V$
plane for $\beta\neq{}0$\@. \      The latter are quite clear in
figure~\ref{Fig03}(b), for example\@. \      Finally, the effect
of relative contraction and elongation of a ruler discussed at the
beginning of this section, which, in general, would mask the
Lorentz contraction, becomes, so to speak, `unmasked' around a
particular point on the ruler moving with any given speed $\beta$
and is easily located\@. \      The Lorentz contraction becomes
visible for a small interval of length $\Delta{}x'$ around this
critical point\@. \\

All of the above observations and deductions follow from our
general formulae obtained, which are easily accessible to students
and sufficiently precise to illustrate faithfully the main
features of the Terrell effect\@. \     The next programme in this
fundamental problem of relativity is how to generalize the
observation site from a pointlike to a non-pointlike one\@. \ This
is a formidable problem which will not be attempted here and
remains to be tackled\@. \\

\end{document}